\documentstyle[twocolumn,eqsecnum,aps,epsf]{revtex} 

\def\abstract#1{\begin{center}{\large Abstract}\end{center} \par #1}
\newtheorem{theorem}     {Theorem}
\newtheorem{lemma}       {Lemma}

\newenvironment{proof}{\noindent{\em Proof.}}%
{{\hspace*{1em}\hfill{$\Box$}}}
{\\}
{{\hspace*{1em} \hfill{$\Box$}}}

\begin{document}
\title{\bf Violation of the cosmic no hair conjecture in 
Einstein-Maxwell-dilaton system} 
\author{Kengo Maeda \thanks{Electronic address: 
g\_maeda@gravity.phys.waseda.ac.jp}, 
Takashi Torii \thanks{Electronic address: torii@th.phys.titech.ac.jp}, 
and Makoto Narita \thanks{Electronic address: narita@se.rikkyo.ac.jp}} 
\address{${}^{\mbox{\rm *}}$
Department of Physics, Waseda University, Oh-kubo, Shinjuku-ku, Tokyo 169, 
Japan}
\address{${}^{\mbox{\rm \dag}}$
Department of Physics, Tokyo Institute of Technology, Oh-Okayama, 
Meguro-ku, Tokyo 152, Japan}
\address{${}^{\mbox{\rm \ddag}}$
Department of Physics, Rikkyo University, Nishi-Ikebukuro, Toshima-ku, 
Tokyo 171, Japan}
\maketitle 
\abstract{The cosmic no hair conjecture is tested in the spherically 
symmetric Einstein-Maxwell-dilaton~(EMD) system with a positive cosmological 
constant $\Lambda$. Firstly, we analytically show that once gravitational 
collapse occurs in the massless dilaton case, the system of field equations 
breaks down inevitably in outer communicating regions or at the boundary 
provided that a future null infinity ${\cal I}^+$ exists. 
Next we find numerically the static black hole solutions in the massive 
dilaton case and investigate their properties for comparison with the 
massless case. 
It is shown that their Abbott-Deser~(AD) mass are infinite, which 
implies that a spacetime with finite AD mass does 
not approach a black hole solution after the gravitational collapse. 
These results suggest that ${\cal I}^+$ cannot appear in the EMD system 
once gravitational collapse occurs and hence the cosmic no hair conjecture 
is violated in both the massless and the massive cases, in contrast to 
general relativity.}\\ 
\section{Introduction}
Recent cosmological observations~\cite{lambda} suggest that there 
must be a positive cosmological constant $\Lambda$ in our universe. 
Furthermore, it is widely believed that the inflation took place in 
the early stage of our universe and the vacuum energy of a scalar 
field plays a role of $\Lambda$ efficiently. 
In such a spacetime, most regions expand exponentially as in de Sitter 
spacetime. However, when the inhomogeneity of the initial matter distribution 
is very high, some regions would collapse into a black hole unless the 
inhomogeneous region is too large~\cite{GarfinkleNakao}, 
or the Abbott-Deser~(AD) mass is negative~\cite{NSM}. 
Such a spacetime is classified as an 
{\it asymptotically de Sitter~(ASD) spacetime}~\cite{SNKM}, 
where there exists a de Sitter-like spacelike future null 
infinity ${\cal I}^+$. 

Gibbons and Hawking~\cite{GibbonsHawking} proposed the cosmic no hair 
conjecture, which states that every spacetime with nonzero $\Lambda$ 
approaches Kerr-Newman-de Sitter spacetime in the stationary limit. 
From a different point of view, the conjecture at least seems to state that 
most regions expand exponentially and the future null 
infinity ${\cal I}^+$ should appear even though gravitational collapse 
occurs somewhere. Although the proof of the conjecture is only shown 
in a restricted class of spacetimes~\cite{BicakP}, it is widely believed that 
any spacetime with $\Lambda$ results in ASD spacetime after gravitational 
collapse in general relativity. 
 
What seems to be lacking, however, is the picture of the 
gravitational collapse in a spacetime with $\Lambda$ in the framework of 
generalized theories of gravity. 
It is important to take some corrections into account when we consider the 
dynamics in spacetime regions with high curvature. 
The purpose of this paper is to investigate the gravitational collapse and 
to test the cosmic no hair conjecture in an effective string theory, which is 
one of promising generalized theories of gravity. Especially, we consider the 
Einstein-Maxwell-dilaton~(EMD) system, which naturally arises from a low 
energy limit of string theory~\cite{Gross}. 

Poletti et al.~\cite{P2} found that the EMD system with 
massless dilaton has no static spherically symmetric black hole solution 
in ASD spacetime. This is essentially due to the fact that there is no regular 
configuration of the dilaton field between a black hole event horizon~(BEH) 
and a cosmological event horizon~(CEH) satisfying the boundary conditions. 
Thus, one may naively 
expect that gravitational collapse cannot occur in the EMD system and hence 
no black holes appear even though highly inhomogeneous region exists. 
However, this is not likely because the EMD system satisfies the dominant 
energy condition, which implies that a highly inhomogeneous region 
would continue to collapse. 
So, what is the final state of the collapse in such a system? 
Motivated by this, we investigate the dynamics of the spherically 
symmetric EMD system when the gravitational collapse occurs. 

The rest of this paper is organized as follows. In Sec.~II 
we write down the basic equations and set the initial conditions. 
In Secs.~III and IV we analytically show that the field equations of the 
EMD system with massless dilaton inevitably break down in the domain of 
outer communicating regions or at the boundary provided that there exists a 
null infinity ${\cal I}^+$. 
Here, we present the detailed proof of our previous result~\cite{MTN}. 
In Sec.~V, we find the static black hole solutions in the EMD system with 
massive dilaton. It is shown that they are stable for linear perturbations 
but their AD mass are infinite. In Sec.~VI we discuss the cosmic 
no hair conjecture in the EMD system on the basis of our results. 

\section{basic equations and initial value conditions}
The action in the EMD system with a positive cosmological 
constant $\Lambda~(>0)$ is 
\begin{eqnarray}
\label{eq-action}
S=\int d^4 x \sqrt{-g}\,
[-R+2(\nabla\phi)^2+2 V_\phi+e^{-2a\phi} F^2+2\Lambda], 
\end{eqnarray}
where $V_\phi$ and $a$ represent the potential and 
the coupling constant of the dilaton field $\phi$, respectively. 
String theory requires $a=1$. 
Varying the action~(\ref{eq-action}), 
we obtain the field equations 
\begin{eqnarray}
\label{eq-pmax}
\nabla_\mu (e^{-2 a \phi}F^{\mu\nu})=0,
\end{eqnarray}
\begin{eqnarray}
\label{eq-pdilaton}
\nabla^2 \phi + \frac{a}{2}e^{-2 a \phi}F^2 
- \frac{1}{2} \frac{\partial V_\phi}{\partial \phi}=0,
\end{eqnarray}
\begin{eqnarray}
\label{eq-pein}
R_{\mu\nu} &=& 2\nabla_\mu \phi \nabla_\nu \phi 
+(V_\phi + \Lambda) g_{\mu\nu} \nonumber \\ 
 &+& 2e^{-2 a \phi}F_{\mu\rho}{F_{\nu}}^\rho-\frac{1}{2}g_{\mu\nu}
e^{-2 a \phi}F^2. 
\end{eqnarray}
In the spherically symmetric system, the Maxwell equation~(\ref{eq-pmax}) 
is automatically satisfied for a purely magnetic Maxwell field 
$F=Q\sin\theta d\theta\wedge d\phi\, (F^2=2Q^2/R^4)$, where 
$Q$ is a magnetic charge and $R$ is a circumference radius.
Because an electrically charged solution is obtained by a duality 
rotation from the magnetically charged one~\cite{GM,GHS}, we have only to 
consider the purely magnetic case. 

\begin{figure}[htbp]
 \centerline{\epsfxsize=5.0cm \epsfbox{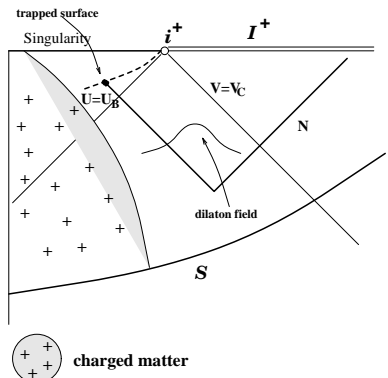}}
         \caption{A Penrose diagram showing the gravitational collapse 
of a charged matter fluid accompanied by a dilaton field supposing that ASD 
spacetime appears. A closed trapped surface is formed as the gravitational 
collapse proceeds. $N$ is a characteristic null hypersurface.} 
               \protect
\label{fig-Ascd8}
\end{figure}

It is worth noting that we cannot take any regular initial 
data $(S,h_{ab},K_{ab})$ in the EMD system because the field 
strength of the Maxwell field $F^2$ diverges at $R=0$, where $h_{ab}$ 
is the metric on a 3-dimensional spacelike hypersurface $S$ embedded 
in $(M,g)$ and $K_{ab}$ is the extrinsic curvature. 
Physically, it seems reasonable to suppose that the existence of 
this central charge results from the gravitational collapse of an 
appropriate charged matter field such as a charged perfect fluid, as 
depicted in Fig.~\ref{fig-Ascd8}. 
In that case, we can take regular initial data $(S,h_{ab},K_{ab})$. 

Once gravitational collapse occurs, a closed trapped surface ${\cal T}$ will 
appear because the matter field falls into a small region. 
As shown in the proof of Ref.~\cite{SNKM}, ${\cal T}$ cannot be seen 
from ${\cal I}^{+}$ and hence a BEH inevitably appears in ASD 
spacetime~\footnote{We should note that ASD spacetime excludes the possibility 
that a naked singularity appears. See Ref.~\cite{SNKM} in detail.}. 
It follows that the ASD spacetime has a null characteristic hypersurface $N$ 
whose boundary is a closed future-trapped surface in the late stage of 
the gravitational collapse~(see Fig.~\ref{fig-Ascd8}). 
For simplicity, we make the physical assumption that all the charged matter 
fields fall into the BEH in the past of $N$. 
In the next section, we will investigate the evolution of the field 
equations~(\ref{eq-pmax})-(\ref{eq-pein}) from $N$ analytically under 
the assumption that spacetime becomes ASD spacetime. 

\section{Analysis of the dynamics of massless dilaton field}
In this section, we analytically investigate the evolution of 
the system from $N$ in the case of massless dilaton field,~i.e. $V_\phi=0$ 
and present a theorem which states that the evolution from any initial 
data on $N$ results in the breakdown of the field equations in 
outer communicating regions or at the boundary in ASD spacetime. 
Since we are interested in effective superstring theory, hereafter we put 
$a=1$. 

This section is divided into two parts. Firstly, we shall 
state some reasonable assumptions and present the theorem. 
As a first step to the proof, we next consider the asymptotic 
behavior of the dilaton field on both event horizons. 

\subsection{Assumptions and Theorem} 
Firstly, we assume that surface gravity of both horizons are 
almost constant asymptotically. By using Gaussian null 
coordinates covering the BEH~(CEH), 
\begin{eqnarray}
\label{eq-gaussian}
ds^2&=& - 2dr\,d\eta + F\,(r,\eta) d\eta^2  \nonumber \\
&+& R^2(r,\eta) (d\theta^2+\sin^2 \theta d\phi^2), 
\end{eqnarray}
let us define the surface gravity of the 
BEH~(CEH) $\kappa_B$~($\kappa_C$) by $F_{,r}/2|_{r=0}$, 
where $\partial_r$ is a future-directed ingoing~(outgoing) 
null geodesic intersecting the BEH~(CEH) and $F=r=0$, $F_{,r}>0$ 
on the BEH~(CEH). 
If $(\partial_\eta)^\mu$ is a timelike killing vector field in $r<0$, 
each surface gravity reduces to the usual surface gravity $\kappa$ 
defined as 
\begin{eqnarray}
\label{eq-surface}
\nabla^{\mu}(\eta^\nu \eta_\nu) = -2\kappa\, \eta^\mu. 
\end{eqnarray}
As shown in~\cite{GibbonsHawking}, each surface gravity and each event horizon 
area corresponds to the temperature and the entropy of the corresponding event 
horizon if the space-time is almost static. 
In addition, it has been shown that the areas of both BEH and CEH 
cannot decrease and have an upper bound, $12\pi/\Lambda$~\cite{SNKM,MKNI}. 
Therefore, it is clear that both areas approach positive constants 
asymptotically. This physically implies that in a neighborhood of 
each horizon, the spacetime would approach locally a static one. 
On the analogy of the almost static spacetime, we shall assume 
\begin{eqnarray}
\label{eq-surfaceB}
\lim_{\eta \to +\infty} \kappa_B \sim \mbox{const.}>0, 
\end{eqnarray}
and 
\begin{eqnarray}
\label{eq-surfaceC}
\lim_{\eta \to +\infty} \kappa_C \sim \mbox{const.}>0, 
\end{eqnarray}
in ASD space-time. The above assumptions indicate 
that if the entropy of the BEH~(CEH) is asymptotically constant, its 
temperature neither goes to $0$~(this is the third law of 
event horizon~\cite{Israel}) nor diverges. 

Next we shall assume that the outgoing null expansion of the CEH 
does not approach $0$ asymptotically. Let us consider Gaussian null 
coordinates covering the CEH and an outgoing null geodesic, $\partial_r$ 
intersecting the CEH. Defining the outgoing null expansion 
$\theta_{+}\equiv 2R_{,r}/R$, the Raychaudhuri 
equation~(see Ref.~\cite{HE}) is 
\begin{eqnarray}
\label{eq-Ray}
\frac{d\theta_{+}}{dr} = -\frac{1}{2}{\theta_{+}}^2-\phi_{,r}^2. 
\end{eqnarray}
This means that $\theta_{+}$ cannot become negative or $0$, otherwise, 
the outgoing null geodesic would not reach the null infinity 
${\cal I}^{+}$~($R=\infty$). Therefore, we also assume 
\begin{eqnarray}
\label{eq-ex-as}
\inf_{\eta} \theta_{C}(\eta)>0, \qquad \eta\in [0,+\infty), 
\end{eqnarray}
or equivalently 
\begin{eqnarray}
\inf_{\eta} R_{,r}(\eta)|_C>0, \qquad \eta\in [0,+\infty), 
\end{eqnarray}
because the area of the CEH, $R_C$ has an upper bound. Here, $\theta_{C}$ 
is the expansion at the CEH. This assumption means that 
$\theta_{+}$ at the CEH cannot approach $0$ in the future. 

Under the above assumptions we will consider the evolution of the 
field equations~(\ref{eq-pdilaton}),~(\ref{eq-pein}) in the massless 
dilaton case. In terms of double null coordinates, 
\begin{eqnarray}
\label{eq-double}
ds^2 &=&-2\,e^{-\lambda}\,(U, V)\,dU dV \nonumber \\
&+& R(U, V)^2 
(d\theta^2+ \sin^2\theta d\phi^2). 
\end{eqnarray}
The field equations~(\ref{eq-pdilaton}),~(\ref{eq-pein}) can be 
reduced to the dynamical field equations 
\begin{eqnarray}
\label{eq-ddi}
2R^3~(R_{,U}\phi_{,V}+R{{\phi}_{,UV}}
+R_{,V}\phi_{,U})
=Q^2 e^{-2\phi-\lambda},
\end{eqnarray}
\begin{eqnarray}
\label{eq-la}
\lambda_{,UV}-\frac{2 R_{,UV}}{R}=
2\phi_{,U}\phi_{,V} + e^{-\lambda}
\left(\frac{Q^2 e^{-2\phi}}{R^4}-\Lambda \right), 
\end{eqnarray}
\begin{eqnarray}
\label{eq-R}
R_{,UV}+\frac{R_{,U}R_{,V}}{R}
=-\frac{e^{-\lambda}}{2R}
\left(1-\frac{Q^2 e^{-2\phi}}{R^2}
-\Lambda R^2 \right),
\end{eqnarray}
and the constraint equations 
\begin{eqnarray}
\label{eq-ru}
{R}_{,UU}+{\lambda}_{,U}{R}_{,U}
=-({\phi}_{,U})^2 R,
\end{eqnarray}
\begin{eqnarray}
\label{eq-rv}
{R}_{,VV}+{\lambda}_{,V}{R}_{,V}
=-({\phi}_{,V})^2 R, 
\end{eqnarray}
where $A_{,X}$ is a partial derivative of $A$ with respect to $X$. 

We present the following theorem. 
\begin{theorem}
\label{theorem}
Let us consider the dynamical evolution of the 
equations~(\ref{eq-ddi})-(\ref{eq-R}) with initial data on the 
characteristic null hypersurface $N$ in ASD spacetime. 
Then, there is a $U_1\,(\le U_B)$ such that the system of equation 
breaks down at $U=U_1$ in the sense that the equations cannot be 
evolved from the hypersurface $N$ to the surface $U_B$, where $U_B$ 
is the coordinate at the BEH. \\
\end{theorem} 

\subsection{Asymptotic behavior of the dilaton field}
We investigate the asymptotic behavior of the dilaton field and give 
the following lemma. 
\begin{lemma}
\label{lemma1}
The asymptotic values of the dilaton field $\phi$ cannot 
diverge at any of the two event horizons. To be precise, 
\begin{eqnarray}
\label{eq-sup}
\sup_\eta |\phi(0,\eta)|<+\infty, \qquad  \eta\in[0,+\infty) 
\end{eqnarray}
must be satisfied in ASD spacetime. 
\end{lemma}

To prove this, let us consider Gaussian null coordinates~(\ref{eq-gaussian}) 
covering the CEH and estimate a quantity $I(\eta)$ defined on each outgoing 
null hypersurface~($\eta=\mbox{const.}$ null hypersurface), 
\begin{eqnarray}
\label{eq-Phimunu}
I(\eta)=\int^{0}_{-r_N(\eta)}\phi_{,r}^2 (r,\eta)\,dr, 
\end{eqnarray}
where $-r_N(\eta)$ is the value of $r$ at the intersection between 
$N$ and $\eta=\mbox{const.}$ null hypersurface. 
Because the area of the CEH has an upper bound, 
$12\pi/\Lambda$, $r_N(\eta)$ also has an upper bound as follows, 
\begin{eqnarray}
\label{eq-affine}   
0<r_N(\eta)&=&\int^{R_C(\eta)}_{R_N(\eta)}\frac{2dR}
{\theta_{+} R}{\Bigg |}_\eta< \int^{R_C(\eta)}_{R_N(\eta)}
\frac{2dR}{\theta_{C} R_N}{\Bigg |}_\eta \nonumber \\
&<& C(R_C(\eta)-R_N(\eta))<C\sqrt{\frac{3}{\Lambda}}, \quad
C(>0),
\end{eqnarray}
where $\theta_C$, $R_C$, and $R_N$ are the values of 
$\theta_+$, $R$ at the CEH and the value of $R$ at $N$, 
respectively~\footnote{We derived the first inequality by using 
the fact that $\theta_{+}$ is a positive and strictly decreasing 
function in the outer regions of the BEH by Eq.~(\ref{eq-Ray}). 
To derive the second inequality, we used the assumption~(\ref{eq-ex-as}).}. 
Therefore, we can obtain 
\begin{eqnarray}
\label{eq-diverge}
\lim_{\eta \to \infty} I(\eta)=+\infty \quad \mbox{when} 
\quad \lim_{\eta \to \infty} \phi(0,\eta)=\pm\infty 
\end{eqnarray}
because the following inequality 
\begin{eqnarray}
\label{eq-estimate1}
2|\phi(0,\eta)-\phi(-r_N,\eta)|&-&r_N(\eta) 
\nonumber \\
&\le& \int^{0}_{-r_N(\eta)}(2|\phi_{,r}|-1)|_\eta\,dr \nonumber \\
&\le& \int^{0}_{-r_N(\eta)}{\phi_{,r}}^2|_\eta dr=I 
\end{eqnarray}
is satisfied and $\phi(-r_N,\eta)$ has a limit at the intersection of 
$N$ and the BEH, 
\begin{eqnarray}
\label{eq-estimate2}
\lim_{\eta \to \infty} \phi(-r_N,\eta)=\phi|_{\mbox{BEH}}. 
\end{eqnarray}
Then, because of the Raychaudhuri equation~(\ref{eq-Ray}), 
$\theta_C$ behaves as 
\begin{eqnarray}
\label{eq-ex-diverge}
\lim_{\eta \to \infty} \theta_C< -\lim_{\eta \to \infty} I= -\infty. 
\end{eqnarray}
This is a contradiction because $\theta_{+}$ must be positive 
in the outer regions of the BEH. 
Similarly, we can also see that $\phi$ on the BEH should not diverge 
asymptotically. Thus, Eq.~(\ref{eq-sup}) must be satisfied at both 
event horizons. $\Box$ \\
Under the above observation, we will prove theorem~\ref{theorem}. 
\section{Proof of the theorem}
We shall prove theorem~\ref{theorem} by contradiction below. In the first 
step, we consider the asymptotic behavior of field functions near 
the CEH. 
It is convenient to rescale the coordinate $U$ in the double null 
coordinates~(\ref{eq-double}) such that $U$ is an affine 
parameter of a null geodesic of the CEH, i.e, $\lambda$ is constant 
along the CEH. 
Hereafter we use the character $u~(=F(U))$ instead of $U$ in this 
parameterization to avoid confusion. 
Under such coordinates, Eq.~(\ref{eq-ru}) on the CEH is 
\begin{eqnarray} 
\label{eq-ruc}
R_{,uu}=-(\phi_{,u})^2 R \le 0.
\end{eqnarray}
By the assumption~(\ref{eq-surfaceC}), $u$ is asymptotically 
related to the coordinate $\eta$ of Eq.~(\ref{eq-gaussian}) by 
\begin{eqnarray}
\label{eq-affi-etau}
u\sim e^{\kappa_C \eta} \quad \mbox{on the CEH}. 
\end{eqnarray}
Therefore, the null geodesic generators of the CEH are future complete 
in asymptotically de Sitter space-time. 
Because the area of the CEH is non-decreasing and has an upper bound, 
\begin{eqnarray}
\label{eq-c-upper}
\lim_{u \to +\infty}R_{,u}=0 \quad \mbox{and} \quad \lim_{u \to +\infty} R=C_1 
\end{eqnarray}
along the CEH, where $C_i~(i=1,2,..)$ is a positive constant. 
This implies that $R_{,uu}(\le 0)$ on the CEH must converge to $0$ faster 
than $u^{-2}$. Hence the future asymptotic behavior of $R_{,uu}$ on the CEH 
is represented as follows, 
\begin{eqnarray}
\label{eq-dR}
\lim_{u \to +\infty} u^2 R_{,uu}=-\lim_{u \to +\infty}f(u)=0, 
\end{eqnarray}
where $f(u)$ is some non-negative function. Then, by Eq.~(\ref{eq-ruc}) 
and lemma~\ref{lemma1} 
\begin{eqnarray}
\label{eq-dph}
\phi_{,u}\sim u^{-1}\sqrt{f}, \qquad \mbox{and} \qquad 
\lim_{u \to +\infty} \phi=\mbox{const.} 
\end{eqnarray}

We shall obtain the asymptotic value of 
$R_{,V}$ on the CEH by solving Eq.~(\ref{eq-R}) as 
\begin{eqnarray}
R_{,V} &=& \left(\int_{u_i}^u \frac{K R}{R_i}du+ R_{,V}|_i \right)
\frac{R_i}{R}, \qquad \\
K &=& -\frac{e^{-\lambda}}{2R}
\left(1-\frac{Q^2e^{-2\phi}}{R^2}-\Lambda R^2 \right), \nonumber 
\end{eqnarray}
where $R_{,V}|_i$ and $R_i$ are initial values of $R_{,V}$ and $R$ 
on $u=u_i$, respectively. $K$ must approach a positive constant 
$K_\infty$ as $u \to +\infty$ 
because $R,\,\phi\to \mbox{const.}$ by lemma~\ref{lemma1} and the expansion 
at the CEH behaves as 
$\lim_{\eta \to +\infty}\theta_{+}\sim \lim_{\eta \to +\infty}u^{-1} 
R_{,V}>0$ by Eq.~(\ref{eq-ex-as}). 
Then, the asymptotic value of $R_{,V}$ is 
\begin{eqnarray}
\label{eq-R'}
R_{,V}\,\sim\,K_\infty u>0. 
\end{eqnarray}

We shall also obtain the asymptotic value of $\phi_{,V}$ on the CEH by 
solving Eq.~(\ref{eq-ddi}) and the solution is the following, 
\begin{eqnarray}
\label{eq-sph'}
\phi_{,V} &=& \left(\int_{u_i}^u \frac{H R}{R_i}du+ \phi_{,V}|_i \right)
\frac{R_i}{R}, \\
H &=& \frac{Q^2 e^{-2\phi-\lambda}}{2R^4}-
\frac{R_{,V}\,\phi_{,u}}{R}, \nonumber 
\end{eqnarray}
where $\phi_{,V}|_i$ is an initial value of $\phi_{,V}$ on $u=u_i$. 
$H$ approaches a positive constant $H_\infty$ as $u \to\infty$ 
because 
$R_{,V}\phi_{,u}\sim \sqrt{f} \to 0$ by Eqs.~(\ref{eq-dph}) 
and (\ref{eq-R'}). 
The asymptotic value of $\phi_{,V}$ is 
\begin{eqnarray}
\label{eq-ph'}
\phi_{,V}\sim H_\infty u>0.
\end{eqnarray}

Let us denote a timelike hypersurface $r=-\epsilon$ by $T_C$, 
where $r$ is the affine parameter of an outgoing null 
geodesic $\partial_r$ in Gaussian null coordinates and 
$\epsilon$ is a small positive constant. If we take $\epsilon$ small 
enough, an infinitesimally small neighborhood~${\cal U}_C$ of 
the CEH contains $T_C$. 
Let us denote each point of the intersection of $T_C$ and 
$u=\mbox{const.}$~hypersurface by $p(u)$. 
By using Eq.~(\ref{eq-ddi}), the solution of $h\equiv\phi_{,u}$ 
along each $u=\mbox{const.}$ is 
\begin{eqnarray}
\label{eq-pu}
h &=& \left(-\int_{V}^{V_C} \frac{LR}{R_C}dV+ h_C \right) \frac{R_C}{R}, \\
L &=& \frac{Q^2 e^{-2\phi-\lambda}}{2R^4}-\frac{R_{,u}\phi_{,V}}{R}, 
\end{eqnarray}
where $R_C$ and $h_C$ are values of $R$ and $h$ on the CEH, $V=V_C$, 
respectively. 
Let us consider the asymptotic value $(R_{,u}\phi_{,V})|_{V_C}$. 
By Eqs.~(\ref{eq-ruc}),~(\ref{eq-dph}), and~(\ref{eq-ph'})
\begin{eqnarray}
\label{eq-ropital}
\lim_{u \to +\infty}R_{,u}\phi_{,V} &\sim& 
\lim_{u \to +\infty} u\int^\infty_u {u'}^{-2}f\,du'  \nonumber \\
 &=& \lim_{u \to +\infty}\frac{u^{-2}f}{u^{-2}}=\lim_{u \to +\infty}f=0, 
\end{eqnarray}
where we used l'Hospital's rule in the first equality. 
Therefore, $L\sim L_\infty>0$ asymptotically. 
Differentiating $h$ by $V$ in Eq.~(\ref{eq-pu}), 
\begin{eqnarray}
\label{eq-hv}
h_{,V}=L+
\frac{R_C R_{,V}}{R^2}
\left(\int_{V}^{V_C} \frac{LR}{R_C}dV
-h_C \right).
\end{eqnarray}
By using Eq.~(\ref{eq-affi-etau}) and 
the relation between Gaussian null 
coordinates~(\ref{eq-gaussian}) and double null 
coordinates~(\ref{eq-double}), we obtain 
\begin{eqnarray}
\label{eq-relation}
\epsilon \sim u\,\delta V \quad \mbox{on the CEH} 
\end{eqnarray}
for large $u$. 
By the relation~(\ref{eq-relation}) for large values of $u$, 
$h_{,V}\sim L_{\infty}+O(\epsilon) + O(\sqrt{f})$ on $T_C$. 
$h_{,V}\sim L_{\infty}>0$ on the $u=\mbox{const.}$ null 
segment $[V|_{p(u)}, V_C]\subset {\cal U}_C$ asymptotically, 
since $\epsilon$ is an arbitrary small value. Here we obtain 
the result in the first step as $h_{,V}\sim L_\infty >0$. 

In the next step we investigate the behavior of $\phi,\phi_{,V}$, and 
$R$ on the BEH, just like in the CEH case. 
We rescale $V$ into $v$ such that $v$ is an affine parameter and 
$\lambda=\mbox{const.}$ on the BEH, while we leave $U$ unchanged. 
The relation between $v$ and $\eta$ of Eq.~(\ref{eq-gaussian}) is 
given by 
\begin{eqnarray}
\label{eq-affi-etav}
v\sim e^{\kappa_B \eta} \quad \mbox{on the BEH}, 
\end{eqnarray}
as given in the CEH case~(\ref{eq-affi-etau}).
Because the area of the BEH is also non-decreasing and it has 
an upper bound, as in the case of the CEH, 
\begin{eqnarray}
\label{eq-upper-BEH}
\lim_{v \to +\infty}R_{,v}=0 \quad \mbox{and} \quad 
\lim_{v \to +\infty} R=C_2. 
\end{eqnarray}
Therefore we have 
\begin{eqnarray}
\label{eq-drv}
\lim_{v \to +\infty} v^2 R_{,vv}=-\lim_{v \to +\infty}g(v) = 0, 
\end{eqnarray}
along the BEH, where $g(v)$ is some non-negative function. 
This means that by Eq.~(\ref{eq-rv}) 
\begin{eqnarray}
\label{eq-phi-BEH}
\phi_{,v}\sim v^{-1}\sqrt{g}, \qquad \mbox{and} \qquad 
\lim_{v \to \infty}\phi=\mbox{const.} 
\end{eqnarray}
at the BEH. By replacing $u$ by $v$ in the argument of the CEH case and 
solving Eqs.~(\ref{eq-ddi}) and (\ref{eq-R}), the asymptotic behaviors of 
$R_{,U}$ and $\phi_{,U}$ become 
\begin{eqnarray}
\label{eq-drv}
R_{,U} \sim v, \qquad \phi_{,U} \sim C_3~v. 
\end{eqnarray}
Hence the first and third terms in the l.h.s. of the dilaton 
field Eq.~(\ref{eq-ddi}) are negligible asymptotically and then 
\begin{eqnarray}
\label{eq-C}
\lim_{v \to +\infty}
k_{,U}=
\lim_{v \to +\infty}
\phi_{,vU}=C_4>0.
\end{eqnarray}
Here we define $k\equiv\phi_{,v}$. Let us consider an 
infinitesimally small neighborhood ${\cal U}_B$ of the BEH. 
There is a small positive $\epsilon$ such that a timelike 
hypersurface $T_B$ with $r=-\epsilon$ is contained in ${\cal U}_B$, 
where $r$ is the affine parameter of an ingoing null geodesic 
$\partial_r$ in Gaussian null coordinates~(\ref{eq-gaussian}). 
By using the relation~(\ref{eq-affi-etav}), we obtain 
\begin{eqnarray}
\label{}
\epsilon\sim v\,\delta U. 
\end{eqnarray}
$k$ on $T_B$ is asymptotically 
\begin{eqnarray}
k|_{T_B} &\sim&
k|_{BEH} + {k_{,U}|_{BEH}}~(-\delta U) \nonumber \\ &\sim& k|_{BEH} - C_4
\,\epsilon\, v^{-1} \nonumber \\ 
&\sim& -(\epsilon~C_4+ \sqrt{g})~v^{-1}.  
\end{eqnarray}
Now, we reached the result in the second step, i.e.,$\phi_{,V}$ is 
negative on $T_B$ for large values $v~(>v_1)$. 

\begin{figure}[htbp]
\centerline{\epsfsize=5.0 cm \epsfbox{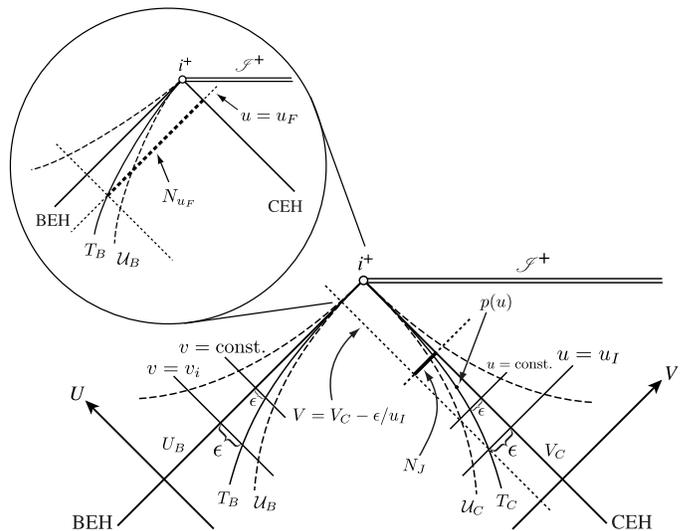}} 
\caption{Timelike hypersurfaces $T_B,\,T_C$ are displayed in the 
neighborhoods ${\cal U}_B$, ${\cal U}_C$ of the BEH and the CEH, respectively. 
A null segment $N_J$ is displayed by a thick line. $\epsilon$ is a 
fixed affine parameter distance along outgoing and 
ingoing null geodesics.}
\protect
\label{fig2-eps}
\end{figure}

Let us consider each $u=\mbox{const}.$ null segment 
$N_{u}(u\ge u_I) :[V_C-\epsilon/u_I,\,V_C]$, 
where $h_{,V}\sim L_\infty >0$ on $N_{u_I}$~(see Fig.~\ref{fig2-eps}). 
If one takes $u_I$ large enough, $N_{u}$ intersects $T_B(v>v_1)$ 
at $u=u_F$. Let us take a sequence of 
$N_{u_J}(J=1,2,...,L+1)$~($L$ is a natural number large enough), 
where $\delta u=(u_F-u_I)/L$ and $u_J=u_I+(J-1)~\delta u$. 
Hereafter we denote $N_{u_J}$ as $N_J$. Before proving theorem~\ref{theorem}, 
we shall establish the following two lemmas. \\
\begin{lemma} 
\label{lemma2}
$R_{,u}<0$ on each $V=\mbox{const.}$ null 
segment~($V_C-\epsilon/u_I \le V< V_C$) between the BEH and $T_C$. \\
\end{lemma}
\begin{proof}
Let us consider Eq.~(\ref{eq-R}). The solution is given by 
\begin{eqnarray}
\label{eq-R,u}
R_{,u}=\left( -\int^{V_C}_V \frac{KR}{R_C} dV + R_{,u}|_{C} \right)
\frac{R_C}{R}.  
\end{eqnarray}
Substituting $V=V_C-\epsilon/u$,  $R_{,u}$ on $T_C$ is asymptotically 
\begin{eqnarray}
\label{eq-R,u1}
R_{,u}(V=V_C-\epsilon/u)
&\sim&  -\epsilon K_{\infty}u^{-1} + R_{,u}|_{C} \nonumber \\
&\sim&   -\epsilon K_{\infty} u^{-1} + O(f)\, u^{-1}. 
\end{eqnarray}
This means that if we take $u_I$ large enough, $R_{,u}<0$ on $T_C$ 
for any $u\ge u_I$ and hence $R_{,u}<0$ on each 
$V=\mbox{const.}$~($V_C-\epsilon/u_I\le V< V_C$) 
null segment between the BEH and $T_C$ 
by Eq.~(\ref{eq-ru}). 
\end{proof}\\
\\
Let us consider a function $Q^2 e^{-2\phi-\lambda}$ on a 
compact region, $V_C-\epsilon/u_I\le V\le V_C$,\,\,$u_I \le u \le u_F$ 
and take the minimum ${Q^2 e^{-2\phi-\lambda}|}_{min}>0$. 
Defining $h_{min}$ by $h_{min}\equiv 
\mbox{min} \{L_{\infty}, {Q^2 e^{-2\phi-\lambda}|}_{min}\}$, 
we give a lemma below. \\
\begin{lemma}
\label{lemma3}
$h_{,V}\ge h_{min} >0$ on $N_{J}$ if $\phi_{,V}>0$ on $N_{J}$. \\
\end{lemma}
\begin{proof}
Let us consider each $u=u_J$ null segment: 
$[V_C-\epsilon/u_J,\,V_C]\in \cal {U_C}$,\, 
$[V_C-\epsilon/u_I,\,V_C-\epsilon/u_J]\notin \cal {U_C}$. 
In the former null segment, $h_{,V}\ge h_{min}>0$, 
as already discussed before. 
In the latter null segment, 
$\int^{V_C}_V (LR/R_C)\, dV \ge 
\int^{V_C}_{V_C-\epsilon/u_J} (LR/R_C)\, dV \sim 
\epsilon/u_J \gg h_C$ 
because $L>0$ by Eq.~(\ref{eq-pu}) and the previous lemma. 
This indicates that $h_{,V}\ge h_{min}>0$ on $N_J$ by Eq.~(\ref{eq-hv}). 
\end{proof}\\
\\
Now, we can prove Theorem~1. 
By the above lemma we can expand 
$\phi_{,V}|_{N_{J+1}}$ by $\delta u$ such that 
$\phi_{,V}|_{N_{J+1}}\cong
(\phi_{,V}|_{N_J}+h_{,V}|_{N_J}\delta u)\ge 
(\phi_{,V}|_{N_J}+h_{min}\,\delta u)$ 
because $\delta u$ is 
arbitrary small and $h_{,V}$ has a positive lower bound, $h_{min}$. 
This means that $\phi_{,V}|_{N_{J+1}}> \phi_{,V}|_{N_{J}}>0$. 
On the other hand, $\phi_{,V}>0$ on $u=u_I$ by Eq.~(\ref{eq-ph'}), 
hence $\phi_{,V}>0$ for each $u_J$ by induction. 
This is a contradiction because 
$\phi_{,V}<0$ on $T_B(u=u_F)$ as shown before. 
Therefore, the assumption that field equations~(\ref{eq-la})-(\ref{eq-rv}) 
could be evolved until $U=U_B$ is false.  \hspace{3cm} $\Box$ \\
\section{Massive Dilaton Case}
In this section, we examine the system with the
massive dilaton field. The unexpected properties in the 
massless dilaton case mainly comes from the fact 
that there is no static black hole solution which the system would 
approach asymptotically after the gravitational collapse\cite{P2}. 
Hence we look for the static solutions in the massive dilaton
case for the first step.

Here, we employ the potential $V(\phi) = 2m_{\phi}^2 e^{2\phi}\phi^2$,
where $m_{\phi}$ is the mass of the dilaton field. We are not sure 
whether this form of the potential is an exact one or not. 
However for small perturbation of the dilaton away from
its vacuum value, we might expect a quadratic form to be a good 
approximation. Moreover, the existence of the static solution
does not depend on the detail of the potential form if it is locally
convex. Actually the results we will show are not changed
qualitatively for 
$V(\phi) = 2m_{\phi}^2 \phi^2$. 
\subsection{static solutions}
We assume the following static chart metric
\begin{eqnarray}
ds^2 &=& -f(t,r) e^{-\delta(t,r)}dt^2  + f(t,r)^{-1} dr^2  \nonumber \\
 & & \;\;\;\;\;\;\;\;\;\;\;\;\;\;\;\;\;\;\;\;\;\;\;\;\;
 + r^2 (d\theta^2+\sin^2\theta d\phi^2),
\end{eqnarray}
\begin{equation}
f(t,r) =1-\frac{2m(t,r)}{r} +\frac{Q^2}{r^2} -\frac{\Lambda}{3}r^2.
\end{equation}
Then the field equations (\ref{eq-pdilaton}) and 
(\ref{eq-pein}) are expressed as
\begin{eqnarray}
-\left[e^{\delta}f^{-1}\dot{\phi}\right]^{\cdot}
  & +&\frac{1}{r^2}\left[r^2e^{-\delta}f\phi' \right]' \nonumber \\
 &&  = e^{-\delta}\left[m_{\phi}^2e^{2\phi}\phi(1+\phi)
   -\frac{e^{-2\phi}Q^2}{r^4} \right],
   \label{dileq}
\end{eqnarray}
\begin{equation}
m' = \frac{r^2}{2}\left[e^{2\delta}f^{-1}\dot{\phi}^2 
   +f \phi^{\prime 2} 
   +m_{\phi}^2e^{2\phi}\phi^2
   +\frac{(e^{-2\phi}-1)Q^2}{r^4}
\right],
   \label{meq}
\end{equation}
\begin{equation}
\delta' = -r\left[e^{2\delta}f^{-2}\dot{\phi}^2 
   +\phi^{\prime 2}
\right] ,
\end{equation}
\begin{equation}
\dot{m} = r^2f\dot{\phi}\phi^{\prime } ,
   \label{beq4}
\end{equation}
where a prime and a dot denote derivatives with respect to 
the radial and time coordinates, respectively.
Here, we have normalized variables and parameters by $\Lambda$ as
$\sqrt{\Lambda} t \to t$, 
$\sqrt{\Lambda} r \to r$, 
$\sqrt{\Lambda} m \to m$, 
$\sqrt{\Lambda} Q \to Q$ and
$\sqrt{\Lambda}m_{\phi}  \to m_{\phi}$.
First we neglect the
terms including time derivative for a while
and look for the nontrivial static solutions.

For the boundary conditions, each functions should be finite in the
$r_{B} \leq r \leq r_{C}$ for regularity.
We can take $\delta(r_{C})=0$ without loss of generality.
If we are interested in some different boundary condition, we can
always have such a boundary condition by only rescaling the
time coordinate.

The equation of the dilaton field (\ref{dileq}) is rewritten as
\begin{eqnarray}
e^{-\delta}f\phi^{\prime \prime }
  &  +&\frac{1}{r^2}\left[r^2e^{-\delta}f \right]' \phi' \nonumber \\
 &&  = e^{-\delta}\left[m_{\phi}^2e^{2\phi}\phi(1+\phi)
   -\frac{e^{-2\phi}Q^2}{r^4} \right].
   \label{horizoneq}
\end{eqnarray}
Since $f=0$ on the  horizons, $\phi'$ is expressed by $\phi$
as
\begin{equation}
 \phi'= \left. \frac1{f'}\left[m_{\phi}^2e^{2\phi}\phi(1+\phi)
   -\frac{e^{-2\phi}Q^2}{r^4} \right]\right|_{\rm horizon}.
\end{equation}
We choose $\phi(r_B)$ and integrate from the BEH to the CEH.
For some value $\phi(r_B)=\phi_1$, the dilaton field
diverges to plus infinity. If we choose larger value
$\phi(r_B)=\phi_2>\phi_1$, it diverges to minus infinity. Then
there is a value $\phi_{\ast} \in (\phi_1, \phi_2)$ with which the 
dilaton field becomes finite everywhere between the
BEH and the CEH. In this sense $\phi(r_B)$ is a shooting parameter
of this system,
which must be determined by the iterative method.

In the massless case, $\phi$ decreases around the BEH because
$f'$ is positive, i.e., the right hand side (r.h.s.) 
in Eq.~(\ref{horizoneq}) 
is negative. Similarly it is locally increasing function
around the CEH because $f'<0$. On the other hand, at the extremum 
of the dilaton field ($\phi'=0$), $\phi^{\prime \prime}<0$
since $f>0$ for $r_B<r<r_C$. Hence the dilaton field does
not have the minimum in this region. This contradict the
behavior around the both horizons. As a result, there
is no spherically symmetric static 
black hole solution in the massless
dilaton case. In the massive dilaton case, however,
the situation is different. As we can see, the mass term
in Eq.~(\ref{horizoneq})  appears in the opposite
sign to the charge term. Hence the sign of the
r.h.s is determined by the value of these two terms.

\begin{figure}[htbp]
\centerline{\epsfxsize=5.0cm \epsfbox{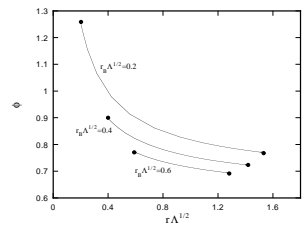}}
\caption{The configurations of the dilaton field of the static black 
hole solutions with $m_{\phi}\sqrt{\Lambda}=0.1$, 
$Q\sqrt{\Lambda}=0.4$ and $r_B\sqrt{\Lambda}=0.2$, $0.4$, 
$0.6$. The left (right) dots correspond to the BEH (CEH).} 

\protect
\label{fig-config}
\end{figure}

Integrating the equations numerically,
we found nontrivial solutions for some
parameters $m_{\phi}$, $Q$ and $r_B$.
We show the configurations of the dilaton field with
$m_{\phi}=0.1$, $Q=0.4$ and several value of $r_B$ in 
Fig.~\ref{fig-config}.
The left (right) end point of each line corresponds to the
BEH (CEH).
The dilaton field decreases monotonically 
between the BEH and the CEH. This implies that
the term including the magnetic charge in Eq.~(\ref{dileq})
is dominant near the BEH while the dilaton mass term
becomes dominant around the CEH. The change of dominant
contribution terms is essential for the existence of the static solutions 
in the massive dilaton case. 

We find solutions with different values of $m_{\phi}$
and $Q$. There are three horizons for the 
Reissner-Nordstr\"{o}m-de Sitter~(RNdS) solution in some parameters, $M$ and 
$Q$. 
Note that since all three horizons degenerate when $Q=0.5$, there is no 
regular RNdS solution for $Q>0.5$. 
From our analysis, new solutions seem not to have inner Cauchy
horizon and we also find the solutions with $Q>0.5$ unlike the 
RNdS case. This is due to the suppression effect of
the magnetic charge by the dilaton field.

\subsection{stability analysis}

Even if the static solutions exist, such objects do not exist
in the physical situation if they are unstable. 
Hence we will investigate the
stability of these new solutions. In this paper, we consider only the radial 
perturbations around the static solutions.

We expand the field functions around the static solution 
$\phi_{0}$, $m_{0}$ and $\delta_{0}$ as 
follows:
\begin{eqnarray}
\phi (t, r) & = & \phi_{0} (r)+ \frac{\phi_1 (t, r)}{r} \epsilon, 
\\
m (t, r) & = & m_{0} (r)+ m_1(t, r) \epsilon, 
\\
\delta (t, r) & = & \delta_{0} (r)+ \delta_1(t, r) \epsilon.
\end{eqnarray}
Here $\epsilon$ is an infinitesimal parameter.
Substituting them into the field functions 
(\ref{dileq})-(\ref{beq4}) and dropping the second
and higher order terms of $\epsilon$,
we find
\begin{eqnarray}
&-& e^{\delta_{0}} f_{0}^{-1} \ddot{\phi}_{1}
+\left[ e^{-\delta_{0}} f_{0}\phi_{1}^{\prime}\right]^{\prime}
 -  \left\{\frac1{r} \left( e^{-\delta_{0}} f_{0}\right)^{\prime}
    \right.
\nonumber
\\   
& & +2 r e^{-\delta_{0}} 
      \left[m_{\phi}^2 e^{2\phi_{0}}\phi_{0}(1+\phi_{0}) 
      -\frac{e^{-2\phi_{0}}Q^2}{r^4}\right] \phi_{0}^{\prime}
\nonumber
\\ 
& & \left. +e^{-\delta_{0}}\left[m_{\phi}^2 e^{2\phi_{0}}(1+2\phi_{0}) 
      +\frac{2e^{-2\phi_{0}}Q^2}{r^4}\right]
   \right\} \phi_{1}  \nonumber \\
& & -\left[\frac2{r} \left( r e^{-\delta_{0}} \phi_{0}^{\prime} 
   \right)^{\prime}
   -2 r e^{-\delta_{0}} \phi_{0}^{\prime 3}\right] m_{1} =0,
\label{leq1}
\end{eqnarray}
\begin{equation}
\dot{m}_1 =  r^2 f_0 \phi_{0}^{\prime} \dot{\phi}_1,
\label{lieq2}
\end{equation}
where $f_{0} = 1-2m_{0}/r+Q^2/r^2-r^2/3$.
Next we set 
\begin{equation}
\phi_{1}  =  \xi(r) e^{i\sigma t}, \;\;\;\;\;
m_{1} =  \eta(r) e^{i \sigma t}.
\end{equation}
If $\sigma$ is real, $\phi$ oscillates around the static
solution and the solution is stable. On the other hand, 
if the imaginary part of $\sigma$ is negative, the
perturbation $\phi_1$ and $m_1$ diverges exponentially 
with time and then the solution is unstable.
By Eq.~(\ref{lieq2}), the relation between $\xi$ and $\eta$  is
\begin{equation}
\eta = r f_{0} \phi_{0}^{\prime} \xi.
\end{equation}
The perturbation equation
of the scalar field becomes
\begin{equation}
-\frac{d^{2}\xi}{dr_{\ast}^2} + U(r) \xi = \sigma^{2} \xi,
\label{leq4}
\end{equation}
where we employ the tortoise coordinate $r_{\ast}$ 
defined by
\begin{equation}
\frac{dr_{\ast}}{dr}  = e^{\delta_{0}}f_{0}^{-1},
\end{equation}
and the potential function is
\begin{eqnarray}
U(r)  = & & e^{-\delta_{0}}f_{0}
  \left\{\frac1r \left( e^{-\delta_{0}} f_{0}\right)^{\prime}
  \right.
\nonumber \\
& &   +2 r e^{-\delta_{0}} 
   \left[m_{\phi}^2 e^{2\phi_{0}}\phi_{0}(1+\phi_{0}) 
      -\frac{e^{-2\phi_{0}}Q^2}{r^4}\right]
   \phi_{0}^{\prime}
\nonumber \\
& &   +e^{-\delta_{0}}\left[m_{\phi}^2 e^{2\phi_{0}}(1+2\phi_{0}) 
      +\frac{2e^{-2\phi_{0}}Q^2}{r^4}\right]
\nonumber \\
& & \left. +   2 f_{0}\phi_{0}^{\prime}
   \left[ \left( r e^{-\delta_{0}} 
   \phi_{0}^{\prime} \right)^{\prime}
   - r^2 e^{-\delta_{0}} \phi_{0}^{\prime 3}
   \right] \right\}.
\end{eqnarray}
Being similar to the other variables, the eigenvalue $\sigma^2$
and the potential function $U$ are normalized as
$\sigma^2/\Lambda \to \sigma^2$ and $U/\Lambda \to U$.

\begin{figure}[htbp]
\centerline{\epsfxsize=5.0cm \epsfbox{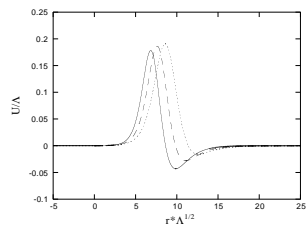}}
\caption{The potential functions of the perturbative equation. 
We set $m_{\phi}\sqrt{\Lambda}=0.1$, 
$r_B\sqrt{\Lambda}=0.5$ and 
$Q\sqrt{\Lambda}=0.2$ (solid line), $0.4$ (dashed line), 
$0.6$ (dotted line).} 

\protect
\label{fig-pot}
\end{figure}

Fig.~\ref{fig-pot} shows the potential functions $U(r)$ of the
solution with $m_{\phi}=0.1$ and $r_B=0.5$.
Since $d^2\xi / dr^{\ast 2}=U(r)
=0$ on both horizons for the negative mode, 
$\xi $ must approach zero as $r^{\ast} \to \pm \infty$
 by the
regularity  of Eq.~(\ref{leq4}). Under this boundary
condition we have searched for the negative eigenmodes.
For the existence of the negative eigenmode, the 
depth of the potential is important. However, the potentials 
in Fig.~\ref{fig-pot} have rather shallow well and we can find no 
negative mode for any $Q$. 
Hence we conclude that the static solutions presented before
are stable at least the radial perturbation level.

\subsection{asymptotic structure}

\begin{figure}[htbp]
\centerline{\epsfxsize=5.0cm \epsfbox{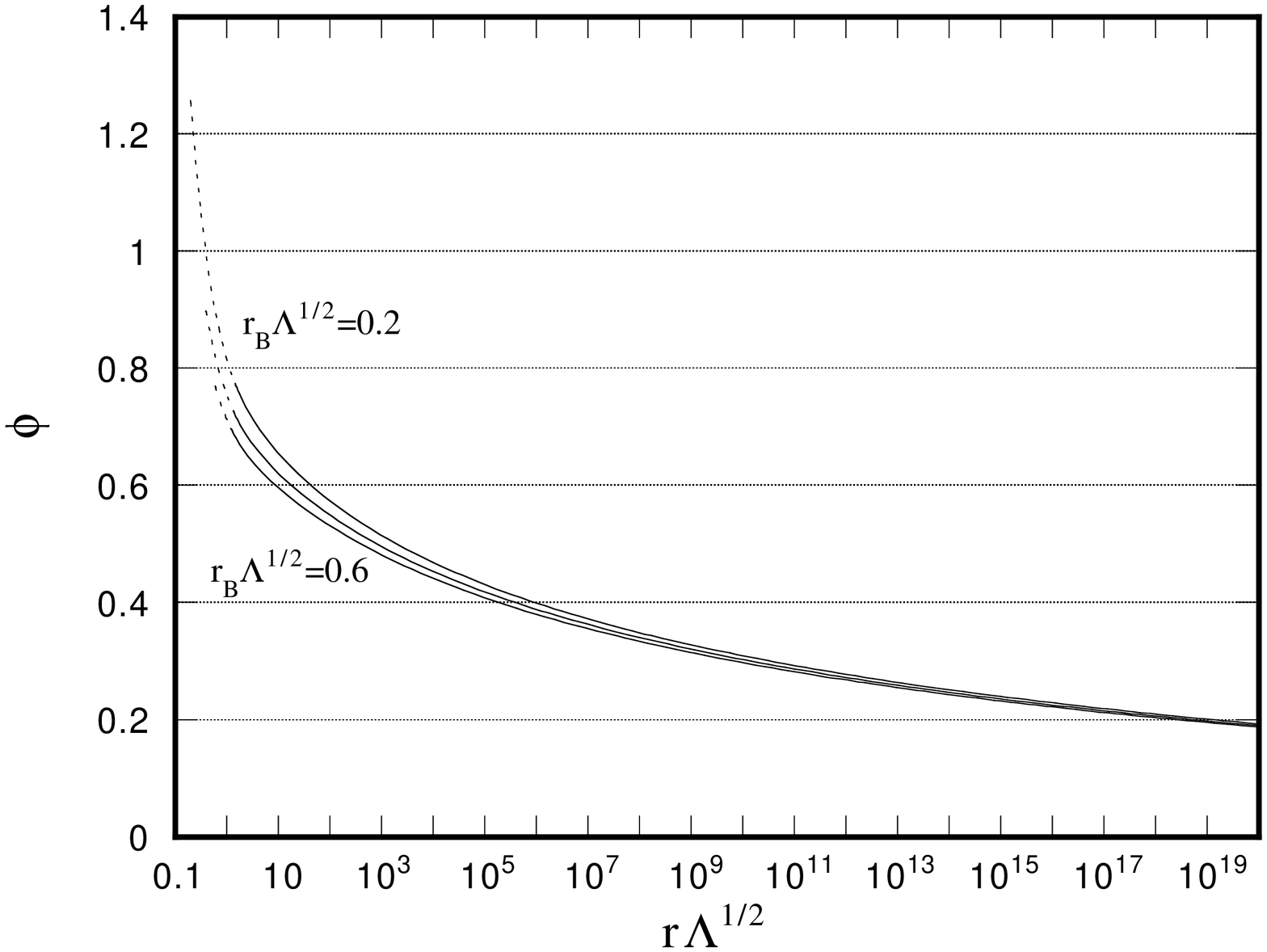}}
\caption{The configurations of the dilaton field of the 
static solutions beyond the CEH. The parameters are 
same as in Fig.~\ref{fig-config}. The dotted part shows 
inside of the CEH. We find that the dilaton fields decays 
extremely slowly.} 

\protect
\label{fig-config2}
\end{figure}

Finally, we will investigate the asymptotic structure of the static black hole 
solutions. 
From the analysis of static solutions, we obtain the boundary 
value of the field functions at the CEH. 
By use of these values we can now
investigate the asymptotic behavior of the field functions
for $r\to \infty$,
which is expected to approach de Sitter spacetime.
Fig.~\ref{fig-config2} shows the field configurations beyond the CEH.
We can find that the dilaton field decreases to its 
potential minimum $\phi =0$. 
Its decay rate is, however, extremely small.
In the asymptotic region, the field equation of the dilaton 
field~(\ref{dileq}) behaves as
\begin{equation}
r^2\phi^{\prime \prime} + 4r \phi' 
    = \frac{m_{\phi}^2}{\lambda} \phi,
\label{asymp}    
\end{equation}
where we assume $f \sim -\lambda r^2$. If the
mass function behaves as $m(r)/r^3\to 0$ as $r\to\infty$,
$\lambda =1/3$.
Putting $\phi \sim r^{-\alpha}$, Eq.~(\ref{asymp}) becomes
\begin{equation}
\alpha (\alpha+1) -4\alpha = \frac{m_{\phi}^2}{\lambda}.
\end{equation}
Hence we find 
\begin{equation}
\alpha  = \frac{3-\sqrt{9-4m_{\phi}^2/\lambda}}{2}.
\end{equation}
By the fact that the decay rate of the dilaton field is very small,
the minus sign was taken. Moreover, in the $m_{\phi}=0.1$
case, we can approximate as
\begin{equation}
\alpha  = \frac{m_{\phi}^2}{3\lambda} \sim 10^{-2}.
\end{equation}
This coincides with the behavior in Fig.~\ref{fig-config2}.
The equation of the mass function (\ref{meq}) behaves as
\begin{eqnarray}
m'  & =  & \frac{r^2}{2}
      \left( -\lambda r^2 \phi^{\prime 2}
      +m_{\phi}^2\phi^{2}\right)  
  \nonumber    \\
   & =  &  \frac{1}{2}
      \left( -\lambda \alpha^2 +m_{\phi}^2\right) r^{2-\alpha}
   \approx \frac{m_{\phi}^2}{2}r^2.
\end{eqnarray}
Hence
\begin{equation}
m\sim  \frac{m_{\phi}^2}{6}r^3.
\end{equation}
This implies that the contribution of the dilaton 
field to the mass function is similar to the cosmological
constant and it diverges as $r \to \infty$. 
Hence the AD mass diverges and this behavior is out of
line with the ASD spacetime\cite{SNKM}.
As a result, the massive dilaton field should also break the asymptotic 
structure similarly to the massless case.

\section{Concluding remarks}
We first tested the cosmic no hair conjecture in an effective string 
theory by investigating the dynamics of the EMD system with massless dilaton. 
We have shown that once gravitational collapse occurs, the system of the 
field equations inevitably breaks down in the domain of outer communicating 
regions or at the boundary under the existence of a de Sitter-like future null 
infinity ${\cal I}^+$. 

In general, the breakdown of the field equations in the EMD system can be 
interpreted as follows: (a) a naked singularity appears in outer communicating 
regions or at the boundary~\cite{comment} or (b) no initial null hypersurface 
$N$ evolves into ${\cal I}^+$. 
For the first case, however, it is hard to understand that a naked 
singularity inevitably appears in any case because the dilaton and the 
electromagnetic fields are essential ingredient for the string 
theory and they are physically reasonable matter fields for testing the cosmic 
censorship conjecture~\cite{Penrose}. 
Thus, we strongly expect that only case (b) is possible and hence the EMD 
system with massless dilaton violates the 
cosmic no hair conjecture~\cite{Private}. 
Here, we should not overlook that our result is independent of the quantity 
of the collapsing mass. 

The above result seems to arise from the fact that the system has no static 
spherically symmetric black hole solution~\cite{P2} and also the dilaton 
field is non-minimally coupled to the electromagnetic field. Let us imagine 
the gravitational collapse in the Einstein-Maxwell~(EM) system for comparison. 
Because all matter fields are minimally coupled to the electromagnetic 
field in the EM system, they fall into a black hole or escape from 
the CEH without difficulty and the resultant spacetime would 
asymptotically approach RNdS one. 
This implies that ASD spacetime appears in the EM system and the 
cosmic no hair conjecture holds, in contrast to the EMD system. 

We next investigated the EMD system with massive dilaton and found that 
the system has a static black hole solution which is stable for 
radial linear perturbations. At first glance, this seems to imply that 
spacetime settles down to the solution after the gravitational collapse 
and the cosmic no hair conjecture holds in the massive dilaton case. 
However, as shown in Sec.~V, the quasi-local mass diverges as $r\to 
\infty$. This means that AD mass also diverges because 
the quasi-local mass corresponds to the AD mass in static spherically 
symmetric spacetime. 
In this sense, this solution is unphysical. In addition, if we consider 
the dynamical evolution from regular initial data with finite AD mass, the 
spacetime seems not to approach our new solutions because AD mass is conserved 
during evolution. This would cause the problem again that the cosmic no hair conjecture is violated in the EMD system, which suggests that the massless 
dilaton case is {\it not} special. 

As a result, we conclude that once gravitational collapse occurs in the 
spherically symmetric EMD system, the spacetime cannot approach ASD spacetime, 
in contrast to the EM system. To the best of our knowledge, this is the 
first counterexample for the cosmic no hair conjecture if the cosmic 
censorship holds. 
It is an open question whether by considering axially symmetric spacetime 
or more general spacetimes one could avoid this problem.

\section*{Acknowledgement}
We express our special thanks to Gary~W.~Gibbons and 
Ted~Jacobson for stimulative discussions and for 
giving us an appropriate interpretation of our results. We are also grateful 
to Shingo Suzuki for helpful discussions and to Akio Hosoya 
and Kei-ichi Maeda for providing us with continuous encouragement. 
We also thanks Gyula Fodor for his careful reading of our manuscript. 
This work is partially supported by Scientific Research 
Fund of the Ministry of Education, Science, Sports, and Culture, by the 
Grant-in-Aid for JSPS~(No. 199704162~(T.~T) and No. 199906147~(K.~M)).

\begin{thebibliography}{99}
\bibitem{lambda}T.~Totani, Y.~Yoshii, and K.~Sato, ApJ {\bf 483}, 
L75~(1997),~S.~Perlmutter, et.al., Nature~{\bf 391}, 51~(1998). 

\bibitem{GarfinkleNakao}D.~Garfinkle and C.~Vuille, Gen.~Relativ.~Gravit. 
{\bf 23}, 471~(1991),~K.~Nakao, Gen.~Relativ.~Gravit~{\bf 24}, 1069~(1992). 

\bibitem{NSM}K.~Nakao, T.~Shiromizu, and K.~Maeda, 
Class.~Quantum.~Grav. {\bf 11}, 2059~(1994). 

\bibitem{SNKM}
T.~Shiromizu, K.~Nakao, H.~Kodama, and K.~Maeda, Phys.~Rev.~D {\bf 47}, 
R3099~(1993). 

\bibitem{GibbonsHawking}G.~W.~Gibbons and S.~W.~Hawking, Phys.~Rev.~D 
{\bf 15},  2738~(1977). 

\bibitem{BicakP}J.~Bicak and J.~Podolsky, Phys.~Rev.~D {\bf 52}, 
887~(1995), J.~Bicak and J.~Podolsky, Phys.~Rev.~D {\bf 55}, 
1985~(1997). 

\bibitem{Gross}D.~Gross and J.~H.~Sloan, Nucl.~Phys. {\bf B~291}, 41~(1987). 

\bibitem{P2}S. J. Poletti, J. Twamley, and D. L. Wiltshire, 
Phys. Rev. D~{\bf 51}, 5720~(1995). 

\bibitem{MTN}
K.~Maeda, T.~Torii, and M.~Narita, 
Phys.~Rev.~Lett. {\bf 81}, 5270~(1998). 

\bibitem{GM}G.~W.~Gibbons and K.~Maeda, Nucl.~Phys. {\bf B~298}, 741~(1988). 

\bibitem{GHS}D.~Garfinkle, G.~T.~Horowitz, and A.~Strominger, 
Phys.~Rev.~D {\bf 43}, 3140~(1991). 

\bibitem{MKNI}
K.~Maeda, T.~Koike, M.~Narita, and A.~Ishibashi, 
Phys.~Rev.~D {\bf 57}, 3503~(1998). 

\bibitem{Israel}W.~Israel, Phys.~Rev.~Lett. {\bf 57}, 397~(1986). 

\bibitem{HE}
S. W. Hawking and G. F. R. Ellis, 
{\it The large scale structure of space time} 
(Cambridge University Press, Cambridge, 1973). 

\bibitem{comment} In the case when a singularity appears along 
the BEH, continuing to the timelike infinity, no naked singularity 
exists in a strict sense~\cite{HE}. However, as mentioned in 
Ref.~\cite{Horowitz}, there exists a spacetime point with very high 
curvature in outer communicating regions near the BEH. 
Thus, we may say that a naked singularity physically occurs in this case. 

\bibitem{Penrose}R.~Penrose, Riv. Nuovo Cimento {\bf 1},~(1969)~252. 

\bibitem{Private} Gary~W.~Gibbons and Ted~Jacobson~(private 
communications). 

\bibitem{Horowitz}G.~T.~Horowitz and S.~F.~Ross, Phys.~Rev.~D. 
{\bf 57}, 1098~(1998). 

\end{thebibliography}
\end{document}